\documentclass[aps,pra,showpacs,twocolumn]{revtex4}

\usepackage{amsmath}
\usepackage{amssymb}  

\newtheorem{Proposition}{Proposition}
\newtheorem{Lemma}{Lemma}

\newcommand{\Tr}{{\textrm{Tr}}}
\newcommand{\Sp}{{\textrm{Sp}}}

\begin{document}

\title{Operations and single-particle interferometry}
\author{Johan \AA berg}
\email[]{johan.aaberg@kvac.uu.se}
\affiliation{Department of Quantum Chemistry, Uppsala University, 
Box 518, SE-751 20 Uppsala, Sweden}
\date{\today}
\begin{abstract}
Interferometry of single particles with internal degrees of freedom is
investigated. We discuss the interference patterns obtained when an
internal state evolution device is inserted into one or both the paths
of the interferometer. The interference pattern obtained is not
uniquely determined by the completely positive maps (CPMs) that
describe how the devices evolve the internal state of a particle. By
using the concept of gluing of CPMs, we investigate the structure of
all possible interference patterns obtainable for given trace
preserving internal state CPMs. We discuss what can be inferred
about the gluing, given a sufficiently rich set of interference
experiments. It is shown that the standard interferometric setup is
limited in its abilities to distinguish different gluings.  A
generalized interferometric setup is introduced with the capacity to
distinguish all gluings.  We also connect to another approach using
the well known fact that channels can be realized using a joint
unitary evolution of the system and an ancillary system. We deduce the
set of all such unitary `representations' and relate the structure of
this set to gluings and interference phenomena.
\end{abstract}

\pacs{03.65.-w, 03.67.-a}

\maketitle

\section{\label{intr}Introduction}

Single-particle interferometry has been widely used to demonstrate
quantum mechanical phenomena. The central question in this
investigation is how interference phenomena are affected when
arbitrary operations are applied to the internal degrees of freedom of
the particle.  Only quite recently has this question received explicit
attention in the literature \cite{Oi, sjomar, peixto}.  These types of
studies are relevant since the transition to general operations
provides a richer structure in the interference
phenomena. Furthermore, general operations may give more realistic
descriptions of interference experiments where noise and decoherence
effects cannot be neglected \cite{matterw}.

It has been shown \cite{Oi} that the interference patterns obtained in
an interferometer are not uniquely determined by the operations
applied. This calls for an investigation of what interference effects
are compatible with a given pair of operations.  By applying the
concept of gluing \cite{ref3, lang} of completely positive maps (CPMs), we
will see that it is the choice of gluing that determines the
interference effects. We are thus able to describe all the interference
effects compatible with given operations.

We also investigate another intuitively reasonable approach to
implement operations in an interferometer, which has been used in
other investigations \cite{Oi, sjomar, peixto}. This uses the well
known fact that operations can be realized using joint unitary
evolution with the system and an ancillary system. Here we investigate
the relation between this approach and the gluing concept in order to
clarify how the choice of joint unitary evolution affects the
interference.

The above questions treat the problem of what interference patterns
are compatible with given channels. We also turn the question around
and ask what information the interference experiments can reveal
about the gluing. It is shown that the ordinary interferometric setup
has only a limited capacity to determine the gluing. However, it is
shown that it is possible to construct a generalized interferometer
for which there is a bijective correspondence between gluings and
interference effects. In Ref.\ \cite{ref3,lang} a complete characterization
of all possible trace preserving gluings of given channels was
developed. The generalized interferometer provides us with a way to
determine these gluings.  As such it opens up for experimental
investigations of these types of problems.

The structure of this article is the following. In Sec.\ \ref{twopath}
the model for the two-path interferometer is introduced. Here we also
make the basic questions of this investigation more precise. In Sec.\
\ref{gluing} the interferometer is discussed in terms of the gluing
concept. By application of the theory developed in Refs.\
\cite{ref1,ref2,ref3,lang}, all possible trace preserving gluings are
expressed. In Sec.\ \ref{deterglue} we deduce all possible
interference effects compatible with given channels. Moreover, we
investigate what can be inferred about an unknown gluing by
performing interference experiments. In Sec.\ \ref{generinfe}, a
generalization of the interferometric setup is introduced. It is shown
that this generalized interferometer has the power to determine
arbitrary unknown trace preserving gluings of two arbitrary known
channels. Section \ref{gluunirep} connects the unitary representation
approach with the gluing approach, by translating results from Refs.\
\cite{ref1,ref2,lang} to the present context. In Sec.\ \ref{allunitary} all
unitary representations of given channels are deduced. The structure
of this set is investigated in terms of gluings, which
makes it possible to select arbitrary gluings of a channel and an
identity channel by a choice of unitary representation. In Sec.\
\ref{discuss} the nature of the nonuniqueness of interference effects 
and gluings is discussed. The conclusions are presented in Sec.\
\ref{conclusions}.

\section{\label{twopath} The two-path interferometer}

The spatial degree of freedom of the interferometer is modeled as a
two-dimensional Hilbert space $\mathcal{H}_{s}$, spanned by
$|1\rangle$ and $|2\rangle$, which correspond to the particle being
localized in paths $1$ and $2$, respectively. The internal Hilbert
space is denoted $\mathcal{H}_{I}$ and the total Hilbert space is
$\mathcal{H}_{s}\otimes\mathcal{H}_{I}$.

The interferometer consists of three parts: First, a `preparation
stage', consisting of a 50-50 beam splitter that creates a
superposition of the particle in the two paths; second, an
`interaction stage', where the state of the particle is
affected; last, the `measurement stage', where a variable phase
shifter is inserted into one of the paths, followed by a second beam
splitter, and finally a detector that determines the presence or
nonpresence of the particle in one of the outgoing paths.

We regard the preparation stage of the interferometer only as a way to
create special types of states on the two paths. If the particle is
initially in path $2$ and the internal state is represented by the
density operator $\rho_{I}$, the first beam splitter creates a state
of the form $\rho_{i}= |\psi\rangle\langle\psi|\otimes \rho_{I}$,
where $|\psi\rangle =\frac{1}{\sqrt{2}}(|1\rangle+|2\rangle)$. This
first beam splitter, as well as the second beam splitter, is modeled
by the unitary operator $U_{bs} = \frac{1}{\sqrt{2}}(|1\rangle\langle
1|+|1\rangle\langle 2|-|2\rangle\langle 1|+|2\rangle\langle 2|)$.

In the interaction stage the total state $\rho_{i}$ of the particle
may change into some new state $\rho_{f}$. This state is thereafter
analyzed in the measurement stage. We return to the interaction stage
below and focus for a moment on the measurement stage.  The phase
shifter is described by the unitary operator $U_{ps} =
|1\rangle\langle 1|+ e^{i\chi} |2\rangle\langle 2|$, where $\chi$ is a
real number. The probability of finding the particle in path $1$,
after the second beam splitter, is \cite{Wagh,Eetal}
\begin{equation}
\begin{split}
p_{1} = & \Tr((|1\rangle\langle 1|\otimes
\widehat{1}_{I})U_{bs}U_{ps}\rho_{f}U_{ps}^{\dagger}U_{bs}^{\dagger})
\\ = & \frac{1}{2} + |E|\cos(\arg(E)-\chi),
\end{split}
\end{equation}
where $E = \langle 1|\Tr_{I}(\rho_{f})|2\rangle$, and where
$\hat{1}_{I}$ is the identity operator on $\mathcal{H}_{I}$.  Thus,
the effect of the measurement stage is to measure the off-diagonal
element of the reduced density operator of the spatial degree of
freedom, in the $\{|1\rangle,|2\rangle\}$ basis. The absolute value
$|E|$ and the argument $\arg(E)$ determine the visibility and the
phase shift, respectively, of the interference pattern.

In the interaction stage some operation acts on the total state of the
particle. Here the words `operation' and `channel' are synonymous with
trace preserving completely positive map \cite{Kraus}.  The operation
is described by a channel $\Phi_{tot}$ that maps the initial total
state $\rho_{i}$ to the final total state $\rho_{f} =
\Phi_{tot}(\rho_{i})$.

Suppose we have a device that can evolve the internal state of a
particle sent through it. The action of this device is described by
the channel $\Phi_{1}$. What is the interference pattern if this
device is inserted into path $1$? One may be tempted to answer that
the interference pattern should be uniquely determined by the channel
$\Phi_{1}$. This is, however, not the case \cite{Oi}. The channel
$\Phi_{1}$ does not provide sufficient information to determine the
interference pattern. The root of this phenomenon is that the total
channel $\Phi_{tot}$ is not uniquely determined by $\Phi_{1}$
\cite{ref3,lang}. One way to put this is to say that the internal state
channel $\Phi_{1}$ is not a `complete' description of the evolution
device when it is to act in a path of an interferometer.  The
following example may clarify the situation.

What is the channel describing a phase shifter? Since the only effect
of the phase shifter is to add an overall phase, it is the identity
CPM. If we prepare particles, let them pass a phase shifter, and then
measure the state of the outgoing particles, the phase shifter has no
measurable effect. However, when inserted into the interferometer, the
effect of the phase shifter is visible as a constant phase shift in
the interference pattern. Hence, the channel describing the phase
shifter, regarded as a device on its own, is not a sufficient
description of the phase shifter when acting inside an interferometer.
 
In this investigation, we wish to find all possible interference
patterns compatible with given internal state evolution channels. We
approach this problem from two different directions.  The first
approach is to note that the total channel $\Phi_{tot}$ can be
regarded as a subspace preserving gluing \cite{ref3,lang} of $\Phi_{1}$
acting in path $1$ and the identity CPM acting in path $2$. We also
consider more general situations with a nontrivial evolution device in
each path of the interferometer. These questions are discussed in
Secs.\ \ref{gluing} - \ref{generinfe}, where we
also discuss what can be inferred about unknown gluings from
interference experiments.

The second approach is to use the well known fact that channels can be
realized using joint unitary evolution of the system and an ancillary
system \cite{Kraus} as \begin{equation}
\label{Ubox}
\Phi_{1}(\rho_{I}) = \Tr_{a}(U_{Ia}\rho_{I}\otimes|a\rangle
\langle a|U_{Ia}^{\dagger}),
\end{equation}  
where $\mathcal{H}_{a}$ is the Hilbert space of the ancillary system
and $|a\rangle$ is a normalized state of the ancilla. A reasonable
method to create an operation $\Phi_{tot}$ would be the
following: Let $U_{sIa}$ be the unitary operator acting on
$\mathcal{H}_{s}\otimes\mathcal{H}_{I}\otimes\mathcal{H}_{a}$ as
\begin{equation}
\label{utot}
U_{sIa} = |1\rangle\langle 1|\otimes U_{Ia}   +  
|2\rangle\langle 2|\otimes\widehat{1}_{I}\otimes\widehat{1}_{a}.
\end{equation}
In words, this means that if the particle passes path $1$ the ancilla
interacts with the particle. If it passes path $2$, then nothing
happens. The total evolution of the particle would then be 
\begin{equation}
\label{globev}
\rho_{f}=\Phi_{tot}(\rho_{i}) = \Tr_{a}(U_{sIa}\rho_{i}\otimes|a\rangle
\langle a|U_{sIa}^{\dagger}).
\end{equation}
If we assume that the initial total state is created with a beam splitter
$\rho_{i} = |\psi\rangle\langle\psi|\otimes\rho_{I}$, the interference
is determined by \cite{sjomar} \begin{equation}
\label{offd2}
E(\rho_{I}) = \frac{1}{2}\Tr(\langle a|U_{Ia}|a\rangle \rho_{I}).
\end{equation}
$E$ is a function from the set of internal state density operators
$\rho_{I}$ to the set of complex numbers. We refer to this function as
the \emph{interference function}.

At first sight this procedure may seem as a straightforward way to
calculate the interference phenomenon caused by a given channel
$\Phi_{1}$. However, the operator $U_{Ia}$, which we use to represent
the internal state evolution device, is not unique. There exist
several unitary operators that realize $\Phi_{1}$ via Eq.\
(\ref{Ubox}). The choice of $U_{Ia}$ affects the interference effect,
as the following example shows.

Suppose we have a channel $\Phi_{1}$ and a representation $U_{Ia}$ of
this channel which gives a nontrivial interference function $E$.
Suppose the internal Hilbert space is of dimension $N<+\infty$.  It
follows that there exists some Kraus representation of $\Phi_{1}$ with
at most $N^{2}$ elements \cite{ref1,lang}. Hence, there exist operators
$V_{k}$ on $\mathcal{H}_{I}$ such that $\Phi_{1}(\rho_{I}) =
\sum_{k=1}^{N^{2}}V_{k}\rho_{I}V_{k}^{\dagger}$.  Assume an ancilla
system with Hilbert space $\mathcal{H}_{a}$ of dimension
$N^{2}+1$. Let $\{|a\rangle, |a_{1}\rangle,\ldots,
|a_{N^{2}}\rangle\}$ be an orthonormal basis of $\mathcal{H}_{a}$. On
$\mathcal{H}_{I}\otimes\mathcal{H}_{a}$ one can construct the
following operator:
\begin{equation}
\label{UKrau}
\begin{split}
U'_{Ia} = &\widehat{1}_{I}\otimes\widehat{1}_{a} -
\widehat{1}_{I}\otimes |a\rangle\langle a| - \sum_{j,l
=1}^{N^{2}}V_{j}V_{l}^{\dagger}\otimes|a_{j}\rangle\langle a_{l}|\\
&+\sum_{j=1}^{N^{2}}V_{j}\otimes |a_{j}\rangle \langle a| +
\sum_{j=1}^{N^{2}}V_{j}^{\dagger}\otimes |a\rangle \langle a_{j}|.
\end{split}
\end{equation}
One can verify that $U'_{Ia}$ is a unitary operator, and also that
$\Phi_{1}$ is obtained if $U'_{Ia}$ is inserted into Eq.\
(\ref{Ubox}), instead of $U_{Ia}$.  Hence, $U_{Ia}$ and $U'_{Ia}$
realize the same CPM $\Phi_{1}$. A global unitary operator $U_{sIa}'$
can be constructed, as in Eq.\ (\ref{utot}), but with $U_{Ia}$
replaced by $U'_{Ia}$. With $U'_{sIa}$, a modified global operation
$\Phi_{tot}'$ can be constructed via Eq.\ (\ref{globev}). For this new
operation the interference function satisfies $E'(\rho_{I}) = 0$ for
every $\rho_{I}$. Hence, there are no interference fringes for any
input state.  In other words, we have constructed two evolution
devices which give the same internal state evolution, but which
nevertheless give rise to two different interference effects. This
example shows that we may choose to set the visibility to zero. In
Sec.\ \ref{allunitary} it is shown that the choice of $U_{Ia}$ may
affect the interference in more general ways. This may be of relevance
for studies like \cite{sjomar}, where the relative phase for CPMs is
defined in terms of unitary representations.

\section{\label{gluing}Gluings}
In this section we introduce the the main tool, gluing of channels,
which we will use to analyze the interferometer. We here give a brief
overview of the concepts developed in Refs.\ \cite{ref1,ref2,ref3,lang},
and translate two results from Ref.\ \cite{ref3,lang}, which will be needed
in the subsequent analysis.

A device, whose effect on the internal state of a particle sent
through it, is described by a channel $\Phi_{1}$. Likewise,
another device is described by a channel $\Phi_{2}$. These devices
are inserted, one in each path of the interferometer. The question is,
what is the 'global' channel $\Phi_{tot}$ that describes the total
operation the single particle has experienced when passing the two
devices? The total Hilbert space of the interferometer can be
decomposed into the orthogonal subspaces
$\Sp\{|1\rangle\}\otimes\mathcal{H}_{I}$ and
$\Sp\{|2\rangle\}\otimes\mathcal{H}_{I}$, each representing pure
states localized in one of the paths. ($\Sp$ denotes the linear span.) 
If the particle is localized in path $1$, then channel $\Phi_{1}$
operates on the internal degree of freedom. If the particle is
localized in path $2$, then channel $\Phi_{2}$ is effected. The set of
all trace preserving gluings of the two channels $\Phi_{1}$ and
$\Phi_{2}$ is precisely the set of all possible total channels
$\Phi_{tot}$ compatible with $\Phi_{1}$ and $\Phi_{2}$
\cite{ref3,lang}. The set of trace preserving gluings of two channels is
the same as the set of \emph{subspace preserving} (SP) gluings of
these two channels \cite{ref3,lang}.

In the present context, the set of SP channels has a rather simple
conceptual interpretation. With respect to the two paths of the
interferometer, a global channel $\Phi_{tot}$ is SP if and only if it
causes no transport of probability weight between the two paths. More
precisely, the channel $\Phi_{tot}$ is subspace preserving if and
only if $\Tr(|1\rangle\langle 1|\otimes
\widehat{1}_{I}\Phi_{tot}(\rho)) =\Tr(|1\rangle\langle 1|\otimes
\widehat{1}_{I}\rho)$ for all density operators $\rho$ on
$\mathcal{H}_{s}\otimes\mathcal{H}_{I}$.

The following proposition is a translation of a general result on SP
gluings \cite{ref3,lang} to the situation considered here.
\begin{Proposition}
\label{allglobal}
Let $\Phi_{1}$ be a channel with linearly independent Kraus
representation $\{V_{n}\}_{n=1}^{N}$ and let $\Phi_{2}$ be a channel
with linearly independent Kraus representation
$\{W_{m}\}_{m=1}^{M}$. All trace preserving gluings of $\Phi_{1}$ and
$\Phi_{2}$ can be written
\begin{equation}
\label{allinter}
\begin{split}
\Phi_{tot}(\rho) = &|1\rangle\langle 1|\otimes
\sum_{n=1}^{N}V_{n}\langle 1|\rho|1\rangle V_{n}^{\dagger} \\ 
& + |2\rangle\langle 2|\otimes\sum_{m=1}^{M}W_{m}\langle
2|\rho|2\rangle W_{m}^{\dagger}\\ &+|1\rangle\langle 2|\otimes
\sum_{n,m}C_{n,m} V_{n}\langle 1|\rho|2\rangle W^{\dagger}_{m}\\ & +
|2\rangle\langle 1|\otimes \sum_{n,m}C_{n,m}^{*} W_{m} \langle
2|\rho|1\rangle V^{\dagger}_{n},
\end{split}
\end{equation}
for all density operators $\rho$ on
$\mathcal{H}_{s}\otimes\mathcal{H}_{I}$, where the matrix $C
=[C_{n,m}]_{n=1,m=1}^{N,M}$ satisfies the condition
\begin{equation}
\label{condisp}
I_{N}\geq CC^{\dagger},
\end{equation}
where $I_{N}$ is the $N\times N$ identity matrix.  Moreover,
 Eq.\ (\ref{allinter}) defines a bijection between the set of trace
preserving gluings and the set of $N\times M$ matrices $C$ that
satisfy Eq.\ (\ref{condisp}).
\end{Proposition}
We will in the following refer to the matrix $C$, of the above
proposition, as the \emph{gluing matrix}. Note that the choice of
linearly independent Kraus representations does not affect the set of
gluings. The Kraus representations play only the role of a
`reference' in terms of which we can describe the gluing using the
gluing matrix. When changing the linearly independent Kraus
representations, the new and the old gluing matrices are related as
$C' = U_{1}CU_{2}^{\dagger}$, where $U_{1}$ and $U_{2}$ are unitary
matrices relating the old Kraus representations to the new ones
\cite{ref1,lang}.

It is to be noted that the above proposition is not formulated
correctly from a technical point of view. It is stated that the CPMs
$\Phi_{1}$ and $\Phi_{2}$ are glued. To be correct we should first
construct a CPM $\overline{\Phi}_{1}$ acting on density operators on
$\Sp\{|1\rangle\}\otimes\mathcal{H}_{I}$. If the original CPM has
Kraus representation $\{V_{n}\}_{n}$, then $\overline{\Phi}_{1}$ has
Kraus representation $\{|1\rangle\langle 1|\otimes
V_{n}\}_{n}$. Similarly one can construct $\overline{\Phi}_{2}$. To be
correct, it is $\overline{\Phi}_{1}$ and $\overline{\Phi}_{2}$ that
are glued. However, since the difference between $\Phi_{1}$ and
$\overline{\Phi}_{1}$ is purely technical, we do not make any
distinction between them here.

The set of channels given by Proposition \ref{allglobal} is rather
`allowing' in the sense that it includes cases where the two devices
may interact or share correlated resources during the operation. If one
wishes to model two \emph{independent} devices, restrictions have to
be imposed on the set of gluings. In Ref.\ \cite{ref2,lang} the concept of
\emph{subspace locality} has been introduced. Subspace locality is
intended to describe a total operation $\Phi_{tot}$ which is composed
from two independent operations, each acting on one location, without
any need for communication or sharing of correlated resources, and
where the two locations are associated with orthogonal subspaces,
rather than a tensor product decomposition.  The following proposition
is a translation of a general result on subspace local gluings
 \cite{ref3,lang}, to the present conditions. The set of subspace
local gluings are called \emph{local subspace preserving} (LSP)
gluings.
\begin{Proposition}
\label{alllocalglobal}
Let $\Phi_{1}$ be a channel with linearly independent Kraus
representation $\{V_{n}\}_{n=1}^{N}$ and let $\Phi_{2}$ be a channel
with linearly independent Kraus representation
$\{W_{m}\}_{m=1}^{M}$. All LSP gluings of $\Phi_{1}$ and
$\Phi_{2}$ can be written
\begin{equation}
\label{islinter}
\begin{split}
\Phi_{tot}(\rho) = &|1\rangle\langle 1|\otimes
\sum_{n=1}^{N} V_{n}\langle 1|\rho|1\rangle V_{n}^{\dagger} \\
 &+ |2\rangle\langle 2|\otimes\sum_{m=1}^{M}W_{m}\langle
 2|\rho|2\rangle W_{m}^{\dagger}\\ &+|1\rangle\langle 2|\otimes V
 \langle 1|\rho|2\rangle W^{\dagger} \\ &+ |2\rangle\langle 1|\otimes
 W\langle 2|\rho|1\rangle V^{\dagger},
\end{split}
\end{equation}
for all density operators $\rho$ on
$\mathcal{H}_{s}\otimes\mathcal{H}_{I}$, where
\begin{equation}
\label{VWdef}
V = \sum_{n=1}^{N} c_{1,n}V_{n},\quad W = \sum_{m =1}^{M}c_{2,m}W_{m},
\end{equation} 
where the vectors $c_{1} = [c_{1,n}]_{n=1}^{N}$ and $c_{2} =
[c_{2,m}]_{m=1}^{M}$ satisfy the conditions
\begin{displaymath}
||c_{1}||^{2} = \sum_{n}|c_{1,n}|^{2} \leq 1,\quad ||c_{2}||^{2} =
\sum_{m}|c_{2,m}|^{2} \leq 1.
\end{displaymath}
 Moreover, if a total channel $\Phi_{tot}$ can be written as above,
 then it is a LSP gluing of $\Phi_{1}$ and $\Phi_{2}$.
\end{Proposition}
Note that the vectors $c_{1}$ and $c_{2}$ are not uniquely determined
by the LSP gluing, but the gluing matrix $C = c_{1}c_{2}^{\dagger}$
is.

The most simple example of a gluing is the gluing of two identity
channels (which is also an example of a LSP gluing). The total CPM is
\begin{equation}
\label{limid}
\begin{split}
\Phi_{tot}(\rho)& = |1\rangle\langle 1|\otimes \langle 1|\rho|1\rangle +
|2\rangle\langle 2|\otimes \langle 2|\rho|2\rangle \\
 &+re^{i\phi}|1\rangle\langle 2|\otimes  \langle 1|\rho|2\rangle + 
re^{-i\phi}|2\rangle\langle 1|\otimes \langle 2|\rho|1\rangle. 
\end{split}
\end{equation}
In this case the gluing matrix is reduced to a single complex number
$c=re^{i\phi}$, with $0\leq r\leq 1$. Although the two channels are
identity channels, there is still a freedom in the choice of
gluing. Suppose the input state of channel (\ref{limid}) is
$\rho_{i}=|\psi\rangle\langle\psi|\otimes\rho_{I}$ with $|\psi\rangle
= \frac{1}{\sqrt{2}}(|1\rangle+|2\rangle)$. The output state is
$\rho_{f} = |1\rangle\langle 1|\otimes\rho_{I}+ |2\rangle\langle
2|\otimes\rho_{I} + re^{i\phi}|1\rangle\langle 2|\otimes\rho_{I} +
re^{-i\phi}|2\rangle\langle 1|\otimes\rho_{I}$. The smaller $r$, the
smaller is the `coherence' between the two paths. Although we have two
identity channels we may nevertheless completely destroy the coherence
by setting $r=0$. Hence, in this case the effect of the gluing is a
relative phase shift and some degree of destruction of coherence
between the two paths.
\section{\label{deterglue}Determining the gluing}
So far we have considered only the structure of the set of gluings on
the two paths of the interferometer. We now turn to the interference
effects caused by these channels. Here we obtain expressions for all
possible interference effects compatible with given
channels. Moreover, we investigate what interference experiments may
tell us about unknown gluings.

To make the analysis as
clear as possible, we assume the input states to be of the form
$\rho_{i} = |\psi\rangle\langle\psi|\otimes\rho_{I}$ with
$|\psi\rangle = \frac{1}{\sqrt{2}}(|1\rangle+|2\rangle)$. This is the
type of states created by the beam splitter, as described in the
Introduction.

Assume channels $\Phi_{1}$ and $\Phi_{2}$ and consider the possible
total channels $\Phi_{tot}$ given by Proposition \ref{allglobal}. One
can deduce the interference function $E$ to be
\begin{equation}
\label{interffunk}
E(\rho_{I}) = \frac{1}{2}\Tr(R\rho_{I}), 
\end{equation}
where
\begin{equation}
\label{SPglop}
R = \sum_{n,m=1}^{N,M} C_{n,m}W_{m}^{\dagger}V_{n},
\end{equation}
with $C_{n,m}$ as in Proposition \ref{allglobal}. In the more
restrictive case of LSP gluings, given by Proposition
\ref{alllocalglobal}, one obtains
\begin{equation}
\label{SPlocp}
R = \sum_{n,m=1}^{N,M} c_{1,n}c_{2,m}^{*}W_{m}^{\dagger}V_{n},
\end{equation}
with the vectors $c_{1}$ and $c_{2}$ as in Proposition
\ref{alllocalglobal}.

We have now  found all the possible interference effects compatible
with two given channels. As seen, all possible choices of interference
effects can be reached by some choice of gluing matrix $C$. As seen
the interference effects are determined by the gluings, not the
channels \emph{per se}.

Since the gluing determines the interference effect this means that the
 interference experiment gives us information about the gluing at
hand. This means that, if we have an unknown gluing, we might possibly
use the interferometer to reveal what gluing we have. In the following
we investigate to what extent this is possible.

Assuming the internal state channels $\Phi_{1}$ and $\Phi_{2}$ are
known, what can be said about the gluing from the interference
experiments? Since the interference function $E$ is linear, it follows
that $E$ is determined by its values on a set of internal states
forming a basis of $\mathcal{L}(\mathcal{H}_{I})$, where
$\mathcal{L}(\mathcal{H}_{I})$ denotes the set of all linear operators
on $\mathcal{H}_{I}$. If $\{|n\rangle\}_{n=1}^{N}$ is an ON-basis of
$\mathcal{H}_{I}$, then the set of density operators
$\{|n\rangle\langle
n|\}_{n}\cup\{|\psi_{nn'}\rangle\langle\psi_{nn'}|,
|\chi_{nn'}\rangle\langle\chi_{nn'}|\}_{n,n':n>
n'}$, where $|\psi_{nn'}\rangle =
\frac{1}{\sqrt{2}}(|n\rangle+|n'\rangle)$, $|\chi_{nn'}\rangle =
\frac{1}{\sqrt{2}}(|n\rangle+i|n'\rangle)$, is such a basis. Given
such a set of interference experiments, the function $E$, and by that
the operator $R$, can be determined. But the task is not to find $R$,
but the gluing matrix $C$. From Eq.\ (\ref{SPglop}) it can be seen
that if $\{W_{m}^{\dagger}V_{n}\}_{n,m=1}^{N,M}$ is a linearly
independent set, then the coefficients $C_{n,m}$ are determined by
$R$. Hence, we can conclude the following.
\begin{Proposition}
\label{eglsp}
Let the CPM $\Phi_{tot}$ be a trace preserving gluing of two channels
with linearly independent Kraus representations $\{V_{n}\}_{n=1}^{N}$
and $\{W_{m}\}_{m=1}^{M}$, respectively. If the set
$\{W_{m}^{\dagger}V_{n}\}_{n,m=1}^{N,M}$ is linearly independent, then
the gluing matrix $C$ is uniquely determined by the interference
function $E$.
\end{Proposition}
It is not always necessary to run the experiment over a basis of
density operators of $\mathcal{L}(\mathcal{H}_{I})$. All information
attainable is extracted for a set of density operators spanning the
subspace $\Sp\{W_{m}^{\dagger}V_{n}\}_{n,m=1}^{N,M}$.  Note also that
Proposition \ref{eglsp} is about the specific type of setup considered
here. As is shown in Sec.\ \ref{generinfe} one can construct
generalized interference experiments that give more information. One
may further note that this proposition gives only a sufficient
condition. It is an open question whether or not it is also a
necessary condition. The condition (\ref{condisp}) may possibly cause
some cases to be uniquely determined in spite of a linearly dependent
set $\{W_{m}^{\dagger}V_{n}\}_{n,m=1}^{N,M}$. Additional constraints,
such as restriction to LSP gluings, may possibly help to determine the
gluing.

The following examples illustrate various situations that may arise.
If one of the devices to be glued is the identity CPM, then $R =
\sum_{m=1}^{M}c_{1,1}c_{2,m}^{*}W_{m}^{\dagger}$.  Since the set
$\{W_{m}\}_{m=1}^{M}$ is linearly independent, it follows that the
gluing matrix (which now is a $1\times M$ matrix) with $C_{1,m} =
c_{1,1}c_{2,m}^{*}$ is uniquely determined. Hence, the gluing, in
sense of the gluing matrix, is uniquely determined. We can conclude
the following.
\begin{Proposition}
\label{glident}
Let $\Phi_{tot}$ be a trace preserving gluing of a channel $\Phi_{1}$
and an identity channel.  The gluing matrix $C$ of $\Phi_{tot}$, with
respect to some linearly independent Kraus representation of the
channel $\Phi_{1}$, is uniquely determined by the interference
function $E$.
\end{Proposition}
Although this is a special case it is a rather important
one. Physically it corresponds to a situation where we have a `black
box' inserted into one of the paths of the interferometer. Using the
interferometer we can investigate evolution caused by this black
box. What Proposition \ref{eglsp} tells us is that the ordinary
interferometer is sufficient to fully explore this black box, with
respect to the gluing property. These aspects will be discussed
further in Sec. \ref{discuss}.

As a second example consider devices $\Phi_{1}$ and $\Phi_{2}$, with
linearly independent Kraus representations $\{|\psi_{1}\rangle\langle
n|\}_{n=1}^{N}$ and $\{|\psi_{2}\rangle\langle m|\}_{m=1}^{N}$,
respectively. Both $|\psi_{1}\rangle$ and $|\psi_{2}\rangle$ are
normalized, and $\{|n\rangle\}_{n=1}^{N}$ is some orthonormal basis of
$\mathcal{H}_{I}$. These two devices have the effect of taking
arbitrary internal states to the pure states
$|\psi_{1}\rangle\langle\psi_{1}|$ and
$|\psi_{2}\rangle\langle\psi_{2}|$, respectively. If
$|\psi_{1}\rangle=|\psi_{2}\rangle$ then $W_{m}^{\dagger}V_{n} =
|m\rangle\langle n|$. The set $\{|m\rangle\langle n|\}_{m,n=1}^{N}$ is
linearly independent and the gluing can be completely determined. If,
on the other hand, the two output states $|\psi_{1}\rangle$ and
$|\psi_{2}\rangle$ are orthogonal, then $W_{m}^{\dagger}V_{n}=0$, and
nothing can be inferred about the gluing. One may note that in this
case the interference function $E$ is identically zero and there are
no interference fringes.

There are cases when it is possible to partially infer the gluing
matrix.  Let both channels $\Phi_{1}$ and $\Phi_{2}$ have the linearly
independent Kraus representation $\{|n\rangle\langle
n|\}_{n=1}^{N}$. This corresponds to devices that set all off-diagonal
elements in the $\{|n\rangle\}_{n=1}^{N}$ basis to zero, but leave
the diagonal elements intact.  One finds that $W^{\dagger}_{m}V_{n} =
\delta_{mn}|n\rangle\langle n|$. Hence, the diagonal elements
$C_{n,n}$ can be determined, but not the off-diagonal elements. This
example also demonstrates that it is not always necessary to run the
interference experiments on a basis of density operators spanning the
whole of $\mathcal{L}(\mathcal{H}_{I})$. Here it is sufficient to run
the experiment for a set spanning the subspace $\Sp\{|n\rangle\langle
n|\}_{n=1}^{N}$. 

With these examples we clearly see that this interferometer cannot
distinguish all gluings. Moreover, we see that its abilities to
recognize the gluings depend on which channels are
glued. Although the interferometer is sufficient in the special case
given by Proposition \ref{glident}, it is problematic as an
experimental tool if one wishes to investigate what gluings of general
type are present in an evolution mechanism.

In the above examples one may recognize a distant analogy with the
problem of an undefined noncyclic geometric phase, because of
vanishing visibility when the interfering states are orthogonal. In
Refs.\ \cite{pistol,offdfisjo,fisjo} the concept of an off-diagonal
geometric phase is introduced, which in some sense extracts more phase
information. In Sec.\ \ref{generinfe} a generalized interferometer is
introduced, which has the ability to completely determine arbitrary
gluings; this seems vaguely analogous to the idea behind the
off-diagonal geometric phase. Note, however, that the geometric phase
is based on given initial states, while  here we consider channels.
\section{\label{generinfe}Generalized interferometry}
It is disturbing that the interferometric setup has only  a limited
capacity to determine the gluing. Here it is shown that there exists a
generalization of the interference setup, with the capacity to
completely determine any trace preserving gluing of any pair of channels.

The standard two-path interferometer determines a detection
probability as a function of a variable phase shift in one of the
paths. This variable phase shift can be regarded as a family of
unitary operators acting on the internal state. This suggests a
generalization, namely, to find the probability as a function of all
unitary operators acting on one of the paths, not only the subfamily
of phase shifts.

In very much the same way as described in Sec.\ \ref{twopath} we consider
a setup with a beam splitter creating an input state $\rho_{i} =
|\psi\rangle\langle\psi|\otimes\rho_{I}$, followed by an interaction
stage with two evolution devices acting according to some gluing.
Then follows a variable unitary operator $U$ in one path, acting on
the total state as $|1\rangle\langle 1|\otimes U +|2\rangle\langle
2|\otimes\hat{1}_{I}$. Finally, there is the second beam splitter and a
measurement of location of the particle. Much as in Sec.\ \ref{twopath},
one finds that the probability of finding the particle in path $1$,
after the final beam splitter, is
\begin{equation}
p_{1}= \frac{1}{2} + |G(U,\rho_{I})|\cos(\arg(G(U,\rho_{I}))),
\end{equation}
\begin{equation}
\label{Gbra}
G(U,\rho_{I}) =
\frac{1}{2}\sum_{n,m=1}^{N,M}C_{n,m}\Tr(W_{m}^{\dagger}UV_{n}\rho_{I}). 
\end{equation}
Although not needed in principle, it may be convenient to add a
variable phase shifter to obtain $p_{1} = \frac{1}{2} +
|G(U,\rho_{I})|\cos(\arg(G(U,\rho_{I}))-\chi)$.  This means that for a
specific choice of $\rho_{I}$ and $U$ one performs ordinary
interference experiments to determine $G(U,\rho_{I})$. We call $G$ the
\emph{generalized interference function}. One may note that
$E(\rho_{I}) = G(\hat{1}_{I},\rho_{I})$.
 
One may wonder if it is not possible to generalize this setup even
further. What if another unitary operator $U'$ is applied to the
second path? Moreover, one may apply unitary operators $\overline{U}$
and $\overline{U}'$ to the two paths \emph{before} the action of the
two evolution devices. However, this does not provide any more
information than does $G$. The generalized interferometer, as
described above, has the power to distinguish all trace preserving
gluings of two known channels.

Lemma \ref{baslemma} and Proposition \ref{GDC} below are formulated in
slightly more general settings than in the rest of this
investigation. Here we allow the internal state channels to have
output on a Hilbert space $\mathcal{H}_{T}$ different from the input
Hilbert space $\mathcal{H}_{S}$. We say that the channels have
\emph{source space} $\mathcal{H}_{S}$ and \emph{target space}
$\mathcal{H}_{T}$ \cite{ref1,lang}. This means that the interferometer
might start with one type of system on the input side, but end in
another type of system on the output side. Propositions
\ref{allglobal} and \ref{alllocalglobal} both remain true under this
generalization, with the modification that the total channel
$\Phi_{tot}$ has source space $\mathcal{H}_{s}\otimes\mathcal{H}_{S}$
and target space $\mathcal{H}_{s}\otimes\mathcal{H}_{T}$. The variable
unitary operator $U$ in the generalized interferometer, as described
above, operates on the target space.
\begin{Lemma}
\label{baslemma}
Let $\{V_{k}\}_{k=1}^{K}$ and $\{W_{k'}\}_{k'=1}^{K}$ be two bases
(not necessarily orthonormal) of
$\mathcal{L}(\mathcal{H}_{S},\mathcal{H}_{T})$. The set of linear
maps $\{\eta_{kk'}\}_{k,k'=1}^{K}$, where the elements are defined as
$\eta_{kk'}(Q) =W_{k'}^{\dagger}QV_{k}$, $\forall
Q\in\mathcal{L}(\mathcal{H}_{T})$, is a basis of
$\mathcal{L}(\mathcal{L}(\mathcal{H}_{T}),\mathcal{L}(\mathcal{H}_{S}))$.
\end{Lemma}
In this lemma $\mathcal{L}(\mathcal{H}_{S},\mathcal{H}_{T})$ denotes
the set of all linear mappings from $\mathcal{H}_{S}$ to
$\mathcal{H}_{T}$.
The proof of this lemma is very similar to a proof in Ref.\
\cite{ref1,lang}. There it is proved that the set
$\{\phi_{kk'}\}_{k,k'=1}^{K}$, defined by
$\phi_{kk'}(Q)=V_{k}QV_{k'}^{\dagger}$, is a basis of
$\mathcal{L}(\mathcal{L}(\mathcal{H}_{S}),\mathcal{L}(\mathcal{H}_{T}))$,
if $\{V_{k}\}_{k}$ is a basis of
$\mathcal{L}(\mathcal{H}_{S},\mathcal{H}_{T})$.
\begin{Proposition}
\label{GDC}
Let the CPM $\Phi_{tot}$ be a trace preserving gluing of two channels
$\Phi_{1}$ and $\Phi_{2}$. The gluing matrix $C$ of $\Phi_{tot}$, with
respect to some linearly independent Kraus representations of the
channels $\Phi_{1}$ and $\Phi_{2}$, is uniquely determined by the
generalized interference function $G$.
\end{Proposition}
The procedure described here can be said to be a process tomography of
the channel $\Phi_{tot}$ \cite{Jones,Tur,Chn,Poy}, but with some \emph{a
priori} information on the process; since we already have the
information on which channels are glued, and wish to determine the
gluing.

\textit{Proof.}
The function $G(U,\rho_{I})$ can be written $G(U,\rho_{I}) =
\frac{1}{2}\Tr(F(U)\rho_{I})$, with $F(U) =
\sum_{n,m=1}^{N,M}C_{n,m}W_{m}^{\dagger}UV_{n}$. For each fixed $U$
the operator $F(U)$ can be determined, given the values of
$G(U,\rho_{I})$ on a set of density operators forming a basis of
$\mathcal{L}(\mathcal{H}_{S})$.

It is always possible to find a basis of
$\mathcal{L}(\mathcal{H}_{T})$ consisting of unitary operators
\cite{Ubas}. Since $F$ is a linear map, it is determined by how it
maps such a basis. Hence, if $G$ is known, the function $F$ is known.

Both $\{V_{n}\}_{n=1}^{N}$ and $\{W_{m}\}_{m=1}^{M}$ are
linearly independent. From these we construct two bases
$\{\widetilde{V}_{n}\}_{n=1}^{K}$ and
$\{\widetilde{W}_{m}\}_{m=1}^{K}$ of
$\mathcal{L}(\mathcal{H}_{S},\mathcal{H}_{T})$, by adding linearly
independent elements. We add these elements in such a way that the first
$N$ ($M$) elements are $\{V_{n}\}_{n=1}^{N}$ ($\{W_{m}\}_{m=1}^{M}$).
The unknown matrix $C$ is extended such that $C_{m,n}= 0$ if $m>M$ or
if $n>N$. With these extensions all the conditions of Lemma
\ref{baslemma} are satisfied.  Hence, the matrix $C$ is uniquely
determined, since it is formed by the expansion coefficients of $F$,
with respect to the basis
$\{\eta_{kk'}\}_{k,k'=1}^{K}$. $\Box$\newline

One can note another approach to constructing an interferometer to
determine the gluing matrix. In this alternative setup the initial
internal state $\rho_{I}$ is fixed, and instead there are two variable
local unitary operators: $\overline{U}$ before, and $U$ after the
evolution devices. This arrangement results in another interference
function $\overline{G}_{\rho_{I}}(U,\overline{U})$. If both the
variable unitary operators act in path $1$, then
$\overline{G}_{\rho_{I}}(U,\overline{U}) =
\sum_{nm}C_{n,m}\Tr(W_{m}^{\dagger}UV_{n}\overline{U}\rho_{I})$. With
an appropriate choice of the initial internal state $\rho_{I}$, the
function $\overline{G}$ can determine arbitrary gluings. By using the
following lemma, which is stated without proof, one can show that
 acceptable initial states have nonsingular density operators.
\begin{Lemma}
\label{densiutilr}
Let $\rho$ be a density operator on $\mathcal{H}$. There exists a set
of unitary operators $\{U_{k}\}_{k=1}^{K}$ such that
$\{U_{k}\rho\}_{k=1}^{K}$ is a basis of $\mathcal{L}(\mathcal{H})$ if
and only if $\rho$ is nonsingular.
\end{Lemma}
One may note that the maximally mixed state is an acceptable choice of
initial state, while a pure state is not.

Although the function $\overline{G}$ or other similar constructions in
principle give the same information as $G$, there may be other aspects
that may make one preferable compared to the others. Apart from the
question of difficulties of experimental realization, there are also
questions about statistics and sensitivity to errors. These questions
are not addressed here, but below we will see another type of
consideration where the choice of setup does matter.

We here relate the material in this and the previous section to some
measures introduced in Ref.\ \cite{Oi}. These measures relate, through
a certain construction, the visibility in an interferometer to Kraus
representations of two given channels inserted into the
interferometer. The dependence on the choice of Kraus representations
one can recognize as the different choices of LSP gluings of the given
channels. That the gluings are LSP can be seen by comparing the
construction in Ref.\ \cite{Oi} with Proposition \ref{LSPunit}, in the
next section. In Ref.\ \cite{Oi} the \emph{coherent fidelity}
$\mathcal{F}_{c}$ between two Kraus representations is defined as the
visibility in the ordinary interferometer, when the initial internal
state is maximally mixed. In the language used here, $\mathcal{F}_{c}$
is the visibility caused by a LSP gluing of the two given
channels. Hence, $\mathcal{F}_{c}(\Phi_{1},\Phi_{2},C) =
2|E(\frac{1}{N}\hat{1}_{I})|$, where $C$ is a LSP gluing matrix $C =
c_{1}c_{2}^{\dagger}$, with respect to some arbitrary choices of
linearly independent Kraus representations. In Ref.\ \cite{Oi} the
maximal coherent fidelity is defined as the the maximum of
$\mathcal{F}_{c}$ over all possible pairs of Kraus representations of
the two channels. This can be recognized as the maximum of
$2|E(\frac{1}{N}\hat{1}_{I})|$ over all possible LSP gluings of
$\Phi_{1}$ and $\Phi_{2}$. In Ref.\ \cite{Oi} it is also determined what is
the closest unitary channel to a given Kraus representation of a
channel. The closest unitary operator is defined as the one giving the
largest visibility for the maximally mixed state as input state, when
the operation acts in one path and the unitary operator acts in the
other path. The maximal visibility so reached can be recognized as the
maximum of $2|G(U,\frac{1}{N}\hat{1}_{I})|$ over all unitary $U$, for a
fixed LSP gluing of the channel and the identity channel.

Using the generalized interferometer one might define several
different measures in the same spirit as in Ref.\ \cite{Oi}. In doing this one
must be aware that the setup may matter in a nontrivial way. We have
seen that the two setups leading to $G$ and $\overline{G}$ are
equivalent in their abilities to determine gluings. However, when
defining measures based on maximizing visibilities, these two setups,
as well as other constructions, may give different answers. As an
example one may consider
$A(\Phi_{1},\Phi_{2},C)=\sup_{U,\rho_{I}}|G(U,\rho_{I})|$, which
corresponds to the maximal visibility over all unitary shifts and
initial internal states. If we restrict to LSP gluings one can deduce
that $A(\Phi_{1},\Phi_{2},C)
=\frac{1}{2}\sup_{||\psi||=1}||V|\psi\rangle||\,\,||W|\psi\rangle||$,
with $V$ and $W$ as in Proposition \ref{islinter}. One may consider
another setup, which is the same as the construction leading to
$\overline{G}$, with the only modification that we also admit
variations of the initial internal state.  The corresponding
interference function is $\widetilde{G}(U,\overline{U},\rho_{I}) =
\frac{1}{2}\sum_{nm}C_{nm}\Tr(W_{m}^{\dagger}UV_{n}\overline{U}\rho_{I})$.
Clearly, knowledge of $\widetilde{G}$ is sufficient to determine the
gluing. In this sense, $\widetilde{G}$ is equivalent to $G$. 
In analogy with the function $A$ one may consider
$B(\Phi_{1},\Phi_{2},C)=
\sup_{U,\overline{U},\rho_{I}}|\widetilde{G}(U,\overline{U},\rho_{I})|$. 
One can show that, in the case of LSP gluings, $B(\Phi_{1},\Phi_{2},C)=
\frac{1}{2}\sup_{||\psi||=1}||V|\psi\rangle||
\sup_{||\chi||=1}||W|\chi\rangle||=
\frac{1}{2}||V||\,\,||W||$. There exist LSP gluings for which
$A(\Phi_{1},\Phi_{2},C) \neq B(\Phi_{1},\Phi_{2},C)$. One example is
if both $\Phi_{1}$ and $\Phi_{2}$ have Kraus representation
$\{|1\rangle\langle 1|,|2\rangle\langle 2|\}$, where
$\{|1\rangle,|2\rangle\}$ is an orthonormal basis of a two-dimensional
$\mathcal{H}_{I}$. We assume that the LSP gluing is such that $V
=|1\rangle\langle 1|$ and $W =|2\rangle\langle2|$. In this case
$A(\Phi_{1},\Phi_{2},C) = \frac{1}{4}$ and $B(\Phi_{1},\Phi_{2},C) =
\frac{1}{2}$. Hence, for these types of questions the choice of
interference setup matters.

\section{\label{gluunirep}Unitary representation of gluings}
In this section we connect the gluing approach with the approach using
unitary channels acting on combinations of the system and ancillary
systems. In Refs.\ \cite{ref1,ref2,lang} it has been shown that for a
special class of CPMs the property of being SP or LSP can be
characterized in terms of unitary actions on system-ancilla
combinations. The present setting of a two-path interferometer belongs
to this special class of CPMs.

The following proposition is a translation of a proposition in Ref.\
 \cite{ref1,lang} to the specific condition considered here.
\begin{Proposition}
\label{SPunit}
A channel $\Phi_{tot}$ is SP on
$(\Sp\{|1\rangle\}\otimes\mathcal{H}_{I},
\Sp\{|2\rangle\}\otimes\mathcal{H}_{I})$
if and only if there exists an ancilla space $\mathcal{H}_{a}$, a
normalized state $|a\rangle\in\mathcal{H}_{a}$, and unitary operators
$U_{1}$ and $U_{2}$ on $\mathcal{H}_{I}\otimes\mathcal{H}_{a}$ such
that
\begin{equation}
\label{unitrep}
\Phi_{tot}(\rho) = \Tr_{a}(U \rho\otimes|a\rangle\langle a| U^{\dagger}),
\end{equation}
for all density operators $\rho$ on
$\mathcal{H}_{s}\otimes\mathcal{H}_{I}$, where
\begin{equation}
U = |1\rangle\langle 1|\otimes U_{1}+|2\rangle\langle 2|\otimes U_{2}.
\end{equation}
\end{Proposition}

Note that every trace preserving gluing of two channels is a SP
gluing. Vice versa, every SP channel is a trace preserving gluing of
two channels \cite{ref3,lang}. The two channels that are glued are
$\Phi_{1}$ and $\Phi_{2}$, which are obtained from $U_{1}$ and
$U_{2}$, respectively, through Eq.\ (\ref{Ubox}). To see this, note
that if the particle is localized in path $1$ with internal state
$\rho_{I}$, then the result of the mapping $\Phi_{tot}$ is again
localized in path $1$, but with the new internal state
$\Phi_{1}(\rho_{I})$.

The following gives a similar construction for LSP channels, and is a
translation of a proposition in Ref.\ \cite{ref2,lang} to the present context.
\begin{Proposition}
\label{LSPunit}
A channel $\Phi_{tot}$ is LSP on
$(\Sp\{|1\rangle\}\otimes\mathcal{H}_{I},
\Sp\{|2\rangle\}\otimes\mathcal{H}_{I})$
if and only if there exist Hilbert spaces $\mathcal{H}_{a1}$,
$\mathcal{H}_{a2}$, normalized vectors
$|a1\rangle\in\mathcal{H}_{a1}$, $|a2\rangle\in\mathcal{H}_{a2}$, a
unitary operator $U_{1}$ on $\mathcal{H}_{I}\otimes\mathcal{H}_{a1}$,
and a unitary operator $U_{2}$ on
$\mathcal{H}_{I}\otimes\mathcal{H}_{a2}$ such that
\begin{equation}
\label{reprU1}
\Phi_{tot}(\rho) = \Tr_{a1,a2}\left(U \rho\otimes|a1\rangle\langle a1|
\otimes|a2\rangle\langle a2|U^{\dagger}\right),
\end{equation}
for all density operators $\rho$ on
$\mathcal{H}_{s}\otimes\mathcal{H}_{I}$, where
\begin{equation}
\label{reprU2}
U = |1\rangle\langle 1|\otimes U_{1}\otimes\widehat{1}_{a2} +
|2\rangle\langle 2|\otimes U_{2}\otimes\widehat{1}_{a1}.
\end{equation}
\end{Proposition}

Comparing Proposition \ref{SPunit} and \ref{LSPunit}, one can see the
difference. For SP gluings the system of interest interacts with one
and the same ancilla system, while for the LSP gluings there are two
ancillary systems. If the particle passes path $1$, it interacts only 
with ancilla $1$, while leaving ancilla $2$ untouched, and the other
way around if the particle passes path $2$. 

In the special case of a gluing of a channel and an identity channel,
Eqs.\ (\ref{reprU1}) and (\ref{reprU2}) are unnecessarily
complicated. In this case all possible gluings, which necessarily are
LSP, can be reached using only one ancillary system. In the next
section we will see that every such gluing can be written as in Eq.\
(\ref{globev}) with a joint unitary operator as in Eq.\ (\ref{utot}),
for a suitably chosen ancillary space.
\section{\label{allunitary}Unitary representation of channels}
As exemplified in the Introduction, one may use a joint unitary
evolution with an ancilla system to implement a channel in one of the
paths of the interferometer. It was also shown that the choice of
unitary representation may affect the interference effects. From
Secs.\ \ref{gluing} and \ref{deterglue} we know that it is the gluing
that determines the interference effects. Moreover, from the previous
section we know that every gluing can be expressed through such
unitary representations. Hence, there must exist some connection
between the choice of unitary representation and the resulting
gluing. The material in the previous sections does not provide us with
any explicit relation between the unitary representations and the
resulting gluing.  Here we establish such a relation, in the special
case of gluings of a channel and an identity channel. Ultimately we
will obtain a strategy to determine which gluing a given unitary
representation gives rise to. Vice versa, if we have a specific gluing
of a channel and an identity channel which we wish to implement, we
will have means to select unitary operators that create precisely this
gluing.  This may be of use in theoretical investigations as well as
in design of actual physical realizations.

We wish to find the relation between the unitary representation of a
channel $\Phi_{1}$ and the LSP gluing $\Phi_{tot}$ which this
representation gives rise to, as described in the Introduction. To do
this we first deduce an expression for the set of all unitary
representations of a given channel $\Phi_{1}$, which then is related
to the LSP gluings. The only limiting assumption is that the Hilbert
space of the internal degree of freedom and the ancillary Hilbert
space are finite-dimensional. The strategy to be used is that every
unitary representation $U_{Ia}$ can be decomposed into two
complementary partial isometries $R$ and $W$, where $R$, say, contains
the `gluing information'. By using this decomposition, an equivalence
relation can be defined on the set of unitary representations, which
tells if these can be distinguished or not in the interferometer. The
equivalence classes correspond to the different LSP gluings.

Let $\Phi_{1}$ be a trace preserving CPM. Let $\mathcal{H}_{a}$ be
finite-dimensional, and let $|a\rangle\in\mathcal{H}_{a}$ be
normalized. We let $\mathbb{U}(\Phi_{1},\mathcal{H}_{a},|a\rangle)$
denote the set of all unitary operators $U_{Ia}$ which represent
$\Phi_{1}$ via Eq.\ (\ref{Ubox}).

The \emph{Kraus number} $K(\Phi_{1})$ of a CPM $\Phi_{1}$
is the number of operators in a linearly independent Kraus
representation of $\Phi_{1}$ \cite{ref1,lang}. One can see that
$\mathbb{U}(\Phi_{1},\mathcal{H}_{a},|a\rangle)$ is empty if
$K(\Phi_{1})>\dim(\mathcal{H}_{a})$.

If an operator $R\in\mathcal{L}(\mathcal{H})$ satisfies
$R^{\dagger}R=P_{i}$ and $RR^{\dagger}=P_{f}$, where $P_{i}$ and
$P_{f}$ are projectors onto two subspaces of $\mathcal{H}$, then $R$
is a \emph{partial isometry} \cite{HS}. We say that the projector
$P_{i}$ projects onto the \emph{initial space} of $R$. Likewise we say
that $P_{f}$ projects onto the \emph{final space} of $R$. One may note
that the subspaces onto which $P_{i}$ and $P_{f}$ project are of the
same dimension. In the following we let $P_{i}^{\perp}$ denote the
complementary projector to $P_{i}$, and similarly with $P_{f}^{\perp}$ and
$P_{f}$.
\begin{Lemma}
\label{partisgru}
Let $\{V_{k}\}_{k=1}^{K}$ be a linearly independent Kraus
representation of a trace preserving CPM $\Phi_{1}$. Let
$\{|a_{1}\rangle,\ldots,|a_{K}\rangle\}$ be an orthonormal set of K
elements in an at least K-dimensional space $\mathcal{H}_{a}$. Then
the operator
\begin{equation}
\label{Rdef}
R = \sum_{k=1}^{K} V_{k}\otimes|a_{k}\rangle\langle a|
\end{equation} 
is a partial isometry. 
\end{Lemma}
To prove this lemma one has to show that
\begin{equation}
\label{PiPfkons}
P_{i} = \hat{1}\otimes |a\rangle\langle a|,\quad P_{f} = \sum_{kk'}
V_{k}V_{k'}^{\dagger}\otimes|a_{k}\rangle\langle a_{k'}|
\end{equation} 
are projectors.
We state without proof the following lemma.
\begin{Lemma}
\label{Udecompo}
Let $U$ be a unitary operator on $\mathcal{H}$. Let $P_{i}$ and
$P_{f}$ be projectors onto two subspaces of equal dimension. If
$P_{f}UP_{i}$ is a partial isometry, then
$P_{f}^{\perp}UP_{i}^{\perp}$ is a partial isometry and
\begin{equation}
\label{eqpar0}
U = P_{f}UP_{i}+P_{f}^{\perp}UP_{i}^{\perp}. 
\end{equation}
\end{Lemma}

Here we introduce some notation.  Let $\mathcal{H}_{a}$ be at least
$K$-dimensional. Let $\mathbb{A}_{K}$ denote the set of all ordered
K-tuples $(|a_{1}\rangle,\ldots,|a_{K}\rangle)$ of pairwise
orthonormal elements in $\mathcal{H}_{a}$. Note that two elements
$\overline{a}, \overline{a}'\in \mathbb{A}_{K}$ are equal if and only
if $|a_{k}\rangle = |a'_{k}\rangle$, $k=1,\ldots,K$.

Let $\{V_{k}\}_{k=1}^{K}$ be a linearly independent Kraus
representation of some channel.  Given $\overline{a}\in
\mathbb{A}_{K}$, let $\mathcal{R}_{\overline{a}}$ denote the range of
the operator $R$ as defined in Eq.\ (\ref{Rdef}). By Lemma
\ref{partisgru}, it follows that $R$ is a partial isometry. 
The initial space of $R$ is $\mathcal{H}\otimes\Sp\{|a\rangle\}$ and
the final space is $\mathcal{R}_{\overline{a}}$.  Let
$\mathbb{W}_{\overline{a}}$ denote the set of partial isometries on
$\mathcal{H}_{I}\otimes\mathcal{H}_{a}$ with initial space
$\left(\mathcal{H}_{I}\otimes\Sp\{|a\rangle\}\right)^{\perp}$ and
final space $\mathcal{R}_{\overline{a}}^{\perp}$.

\begin{Proposition}
\label{Ukonstr}
Let $\{V_{k}\}_{k=1}^{K}$ with $K = K(\Phi_{1})$ be a linearly
independent Kraus representation of the channel $\Phi_{1}$. Let
$\mathcal{H}_{a}$ be at least $K$-dimensional and let
$|a\rangle\in\mathcal{H}_{a}$ be normalized. Then
\begin{equation}
\label{bij}
U_{Ia} = W + \sum_{k=1}^{K} V_{k}\otimes|a_{k}\rangle\langle a|
\end{equation}
 defines a bijection between the set
 $\mathbb{U}(\Phi_{1},\mathcal{H}_{a},|a\rangle)$ and the set of all
 pairs $(\overline{a},W)$ with $\overline{a}\in \mathbb{A}_{K}$ and
 $W\in \mathbb{W}_{\overline{a}}$.
\end{Proposition}
\textit{Proof.}
First it is proved that if $\overline{a}\in \mathbb{A}_{K}$ and $W\in
\mathbb{W}_{\overline{a}}$ then the operator $U_{Ia}$ defined by Eq.\
(\ref{bij}) belongs to
$\mathbb{U}(\Phi_{1},\mathcal{H}_{a},|a\rangle)$. One can verify that
$U_{Ia}$, so defined, is unitary since it is a sum of two
complementary partial isometries. Moreover, one can verify that
$U_{Ia}$ represents $\Phi_{1}$ via Eq.\ (\ref{Ubox}).  Hence,
$U_{Ia}\in \mathbb{U}(\Phi_{1},\mathcal{H}_{a},|a\rangle)$.

It has to be shown that if $U_{Ia}\in
\mathbb{U}(\Phi_{1},\mathcal{H}_{a},|a\rangle)$ then there exist
$\overline{a}\in \mathbb{A}_{K}$ and $W\in \mathbb{W}_{\overline{a}}$,
which give $U_{Ia}$ via Eq.\ (\ref{bij}). Let
$\{|b_{l}\rangle\}_{l=1}^{N}$ be an arbitrary orthonormal basis of
$\mathcal{H}_{a}$.  It follows that $\{W_{l}\}_{l=1}^{N}$ with $W_{l}
= \langle b_{l}|U_{Ia}|a\rangle$ is a Kraus representation of
$\Phi_{1}$. Let $\{V_{k}\}_{k=1}^{K}$ be a linearly independent Kraus
representation of $\Phi_{1}$. It is well known \cite{Preskill} that
any two Kraus representations can be connected through a unitary
matrix, where the Kraus representation with the smaller number of
elements is padded with zero-operators in such a way that the two sets
have the same number of elements. Note that the set $\{W_{l}\}_{l=1}^{N}$
has at least as many elements as $\{V_{k}\}_{K=1}^{K}$, since the last
is a linearly independent Kraus representation \cite{ref1,lang}. The
existence of a unitary matrix connecting padded sets of Kraus
operators is equivalent to the existence of an $N\times K$ matrix $M$
such that $W_{l}= \sum_{k=1}^{K}M_{lk}V_{k}$, for all $l=1,\ldots,N$,
and such that $M^{\dagger}M = I_{K}$, where $I_{K}$ denotes the
$K\times K$ identity matrix. Define the set
$\{|a_{k}\rangle\}_{k=1}^{K}$ by $|a_{k}\rangle =
\sum_{l=1}^{N}M_{lk}|b_{l}\rangle$ for $k=1,\ldots,K$. One can verify
that $\{|a_{k}\rangle\}_{k=1}^{K}$ is an orthonormal set. Let $P_{i}$
and $P_{f}$ be defined as in Eq.\ (\ref{PiPfkons}). Using the fact
that $\Phi_{1}$ is trace preserving, one can verify that
$P_{f}U_{Ia}P_{i}= R$, with $R$ defined as in Eq.\ (\ref{Rdef}). Since
$\{|a_{k}\rangle\}_{k=1}^{K}$ and $\{V_{k}\}_{k=1}^{K}$ satisfy the
properties required by Lemma \ref{partisgru}, it follows that $R$ is a
partial isometry. By Lemma \ref{Udecompo} it follows that $U_{Ia}$ can
be written as in Eq.\ (\ref{bij}) with $W =
P_{f}^{\perp}U_{Ia}P_{i}^{\perp}$. By Lemma \ref{Udecompo} it follows
that $W$ is a partial isometry with the correct initial and final
spaces.
 
Finally, it has to be shown that if two pairs $(\overline{a},W)$ and
$(\overline{a}',W')$ are different, then the corresponding operators
$U_{Ia}$ and $U'_{Ia}$ are different. Assume these two pairs are
mapped to the same $U$. Then
\begin{equation}
\label{simul1}
W-W' = \sum_{k=1}^{K} V_{k}\otimes(|a'_{k}\rangle-|a_{k}\rangle)\langle a|.
\end{equation}
The operator $W-W'$ maps elements in
$\mathcal{H}_{I}\otimes\Sp\{|a\rangle\}$ to the
zero element. Similarly, $\sum_{k=1}^{K}
V_{k}\otimes(|a'_{k}\rangle-|a_{k}\rangle)\langle a|$ maps elements in
$\left(\mathcal{H}_{I}\otimes\Sp\{|a\rangle\}\right)^{\perp}$ to the
zero element. Hence, from Eq.\ (\ref{simul1}) it follows that $W-W' =
0$ and $\sum_{k=1}^{K}
V_{k}\otimes(|a'_{k}\rangle-|a_{k}\rangle)\langle a| = 0$. Let
$|\chi\rangle\in\mathcal{H}_{a}$ be arbitrary. By applying
$\langle\chi|$ `from the left' and $|a\rangle$ `from the right' onto
the last expression, one obtains
$\sum_{k=1}^{K}(\langle\chi|a'_{k}\rangle-\langle\chi|a_{k}\rangle)
V_{k} = 0$. By linear independence of $\{V_{k}\}_{k=1}^{K}$, and the
arbitrariness of $|\chi\rangle$ it follows that
$\overline{a}'=\overline{a}$. Hence, no two distinct pairs are mapped
to the same unitary operator.  $\Box$\newline

Using the interferometric setup, as described in the Introduction, two
unitary representations $U_{Ia}$ and $U'_{Ia}$ are distinguishable in
the interferometer, if and only if the corresponding interference
functions $E$ and $E'$ are different. From Eq.\ (\ref{offd2}) and
Proposition \ref{Ukonstr} it follows that the interference function is
$E(\rho_{I}) = \frac{1}{2}\sum_{k=1}^{K}\langle a_{k}|a\rangle
\Tr(V_{k}\rho_{I})$. Because of the linear independence of
$\{V_{k}\}_{k=1}^{K}$, two unitary representations are
distinguishable if and only if the corresponding vectors $(\langle
a_{k}|a\rangle)_{k=1}^{K}$ and $(\langle a'_{k}|a\rangle)_{k=1}^{K}$
are different.  Since $\{V_{k}\}_{k=1}^{K}$ is a linearly independent
Kraus representation of $\Phi_{1}$, it follows that $(\langle
a_{k}|a\rangle)_{k=1}^{K}$ can be identified with the $1\times
K(\Phi_{1})$ gluing matrix $C$.  As shown in Sec.\ \ref{deterglue},
the gluing matrix is uniquely determined by the interference function
$E$, for this type of gluing. From this it follows that two unitary
representations $U_{Ia}$ and $U'_{Ia}$ are distinguishable in the
interferometer if and only if they correspond to different LSP gluings
of the channel $\Phi_{1}$ and the identity channel. Another way to put
this is to say that $\mathbb{U}(\Phi_{1},\mathcal{H}_{a},|a\rangle)$
can be equipped with an equivalence relation $\sim$. Two unitary
representations are equivalent, $U_{Ia}\sim U'_{Ia}$, if $(\langle
a_{k}|a\rangle)_{k=1}^{K}=(\langle a'_{k}|a\rangle)_{k=1}^{K}$. As we
have seen, this is equivalent to being indistinguishable by the
interferometer, which is the same as saying that they correspond to
the same LSP gluing of $\Phi_{1}$ and the identity channel.

Since the gluing matrix $C$, in the present case, is only a row (or
column) matrix, it can be regarded as a vector.  If this vector $C$
satisfies $||C|| =1$, we say that the LSP gluing is \emph{maximal}.

Using Eqs.\ (\ref{utot}) and (\ref{globev}), it is possible to define
a mapping $\mathcal{M}$ from the set
$\mathbb{U}(\Phi_{1},\mathcal{H}_{a},|a\rangle)$ to the set of LSP
gluings of the channel $\Phi_{1}$ and the identity channel.
\begin{Proposition}
\label{ancill}
Let $\Phi_{1}$ be a channel. Let $\mathcal{H}_{a}$ be
finite-dimensional and $|a\rangle\in\mathcal{H}_{a}$ normalized.
\begin{itemize}
\item  If $\dim(\mathcal{H}_{a})<K(\Phi_{1})$ then 
 $\mathbb{U}(\Phi_{1},\mathcal{H}_{a},|a\rangle)$ is empty.
\item If $\dim(\mathcal{H}_{a})=K(\Phi_{1})$ then 
$\mathcal{M}$ defines a bijection between the set of equivalence
classes under $\sim$ and the set of maximal LSP gluings of $\Phi_{1}$
and the identity channel.
\item If $\dim(\mathcal{H}_{a})>K(\Phi_{1})$ then 
$\mathcal{M}$ defines a bijection between the set of equivalence
classes under $\sim$ and the set of LSP gluings of $\Phi_{1}$ and the
identity channel.
\end{itemize}
\end{Proposition}
In essence this proposition says that if the dimension of the Hilbert
space of the ancilla is equal to the Kraus number of the channel
$\Phi_{1}$, then we reach precisely the maximal LSP gluings through
the unitary representations. If the dimension of the ancillary Hilbert
space is strictly larger than the Kraus number, then we reach all LSP
gluings of $\Phi_{1}$ and the identity channel.
  
\textit{Proof.}
The first statement follows since $K(\Phi_{1})$ is the minimal number
of elements in any Kraus representation of $\Phi_{1}$ \cite{ref1,lang}.

For the second statement, assume
$U_{Ia}\in\mathbb{U}(\Phi_{1},\mathcal{H}_{a},|a\rangle)$. By using
Proposition \ref{Ukonstr} and Eqs.\ (\ref{utot}) and (\ref{globev}),
one finds that the gluing matrix is $C = [\langle
a_{k}|a\rangle]_{k=1}^{K}$. Since $\mathcal{H}_{a}$ is K-dimensional,
it follows that $\{|a_{k}\rangle\}_{k=1}^{K}$ is an orthonormal basis
of $\mathcal{H}_{a}$. Since $|a\rangle$ is normalized,
$\sum_{k}|c_{k}|^{2} = \sum_{k=1}^{K}|\langle a_{k}|a\rangle|^{2} =
1$. Hence, the gluing is maximal. One can see that all elements in an
equivalence class are mapped to the same gluing. Moreover, two
elements from different equivalence classes are mapped to different
gluings.

It has to be shown that every maximal gluing can be reached via
$\mathcal{M}$. Suppose we have a maximal gluing with gluing matrix
$C$. If we regard the gluing matrix as a vector, it follows that $C\in
\mathbb{C}^{K}$, such that $||C||=1$. It is always possible to find an
orthonormal basis $\overline{a}=\{|a_{k}\rangle\}_{k=1}^{K}$ of
$\mathcal{H}_{a}$, such that $C_{k} = \langle a_{k}|a\rangle$.  Let
$U_{Ia}$ be defined from $\{|a_{k}\rangle\}_{k=1}^{K}$, through Eq.\
(\ref{bij}), for some arbitrary choice of $W\in
\mathbb{W}_{\overline{a}}$.

For the case $\dim(\mathcal{H}_{a})>K(\Phi_{1})$ one can reason very
similarly as above, with the modification that
$\{|a_{k}\rangle\}_{k=1}^{K}$ spans a proper subspace of
$\mathcal{H}_{a}$. This implies that the gluing matrix does not have
to be maximal, and we can reach all the LSP gluings.  $\Box$\newline

\section{\label{discuss}Discussion}
As we have demonstrated it is not the internal state channels \emph{per se}
that determine the interference pattern, but their gluings. Even if it
assumed that the devices are acting independently of each other (LSP
gluing), there remains an arbitrariness in the interference pattern,
which corresponds to the nonuniqueness of LSP gluings. Here we
concentrate on the special case of gluings of a channel $\Phi_{1}$ and
an identity channel. Such gluings can be described as pairs
$(\Phi_{1},V)$, where $V$ is as in Eq.\ (\ref{VWdef}). One way to
understand the nonuniqueness is to describe the state of the particle
in the interferometer in terms of an occupation number
representation. This describes the occupation states of the two paths,
rather than the location of the particle. It is sufficient to extend
the Hilbert space of the internal degree of freedom with one
additional dimension spanned by a `vacuum state', which describes the
nonpresence of the particle in that path \cite{ref2,lang}. The total
extended Hilbert space is the tensor product of two such extended
Hilbert spaces. In the case of a trace preserving gluing of a channel
and an identity channel, the corresponding channel in the occupation
number representation can be written as a product channel
$\widetilde{\Phi}_{1}\otimes\widetilde{I}_{2}$, where
$\widetilde{I}_{2}$ is the identity channel acting on operators on the
extended Hilbert space of the empty path. The channel
$\widetilde{\Phi}_{1}$ takes the form \cite{ref2,lang}
\begin{equation}
\label{extended}
\widetilde{\Phi}_{1}(\widetilde{\rho}) = 
\sum_{k}V_{k}\widetilde{\rho}V_{k}^{\dagger} + 
V\widetilde{\rho}|0\rangle\langle 0| +
|0\rangle\langle 0|\widetilde{\rho}V^{\dagger} + 
|0\rangle\langle 0|\widetilde{\rho}|0\rangle\langle 0|,
\end{equation}
where $|0\rangle$ is the vacuum state of the path in which
$\widetilde{\Phi}_{1}$ acts.  As seen, the extended channel
$\widetilde{\Phi}_{1}$ contains the same information as the pair
$(\Phi_{1},V)$. To every trace preserving gluing of the channel
$\Phi_{1}$ and the identity channel, there corresponds a channel
$\widetilde{\Phi}_{1}$. The channel $\widetilde{\Phi}_{1}$ describes not only
 what the machine does with a particle present in the input,
but also what is does with superpositions of the particle and the
vacuum state. For more details concerning this occupation number
approach the reader is referred to Refs.\ \cite{ref2,ref3,lang}.

In Sec.\ \ref{deterglue} (Proposition \ref{glident}) we saw that
the ordinary interferometric setup has the power to determine trace
preserving gluings of a channel and an identity channel. Hence, it can
determine the operator $V$ in Eq.\ (\ref{extended}). In other words, the
interferometer has the capacity to reveal more about the global
evolution than direct measurements as pointed out in Ref.\
\cite{Oi}. However, the equivalent description in terms of
$\widetilde{\Phi}_{1}$ suggests that another strategy is possible, at
least in principle. If the evolution device is subjected to a process
tomography on the
\emph{extended} Hilbert space, the channel $\widetilde{\Phi}_{1}$
would be revealed and hence provide the same information as the
interference experiments would. This would correspond to preparing
states including linear combinations of the particle in some internal
state and the vacuum state. Similarly, the measurements performed on
the output has to be sufficiently rich on the extended state
space. Leaving aside the question of how such states actually would be
produced, and how such measurements would be performed, this means
that the interferometer is not really necessary to determine trace
preserving gluings of a channel and the identity channel. The same
information could, in principle, be obtained with direct measurement
on the output states, provided the input states and the measurements
are sufficiently general on the extended Hilbert space.
\section{\label{conclusions}Conclusions}
Two-path single-particle interferometry of particles with an internal
degree of freedom is investigated. Given internal state evolution
devices, whose action are characterized by trace preserving completely
positive maps (channels), we ask how the interference phenomena are
affected when such devices are inserted into the paths of the
interferometer. We investigate the nonuniqueness of the interference
patterns for given internal state evolution channels. This question is
approached from two points of view. The first is to use the concept of
gluing of completely positive maps developed in Ref.\ \cite{ref3,lang}. It
is found that the possible interference effects are determined by the
gluings, rather than the internal state channels \emph{per se}. Using the
gluing approach we deduce all possible interference effects compatible
with given channels.

In the second approach we make use of the fact that channels can be
realized using joint unitary evolution on a system and an ancillary
system. By this approach we connect to other investigations in the
literature \cite{Oi, sjomar, peixto} in which joint unitary evolution
is used in interferometers.  The choice of joint unitary evolution
used to realize a given channel is not unique. Although two such
unitary operators realize the same channel, they may cause different
interference phenomena when the machine is inserted into one of the
paths of the interferometer. We investigate which gluing each choice of
unitary representation gives rise to, and hence which interference
pattern. Conversely, if one wishes to construct a specific gluing we
determine the possible choices of unitary representations which
give the desired gluing. This may be of use in the design of actual
physical implementations of this type of channel.

In previous work \cite{ref3,lang} the set of all possible trace preserving
gluings of given pairs of channels has been deduced. Here we extend
this work by investigating how interferometers can be used to analyze
which gluing is actually present.  It is shown that the standard
interferometer in general has a limited capacity to determine the
gluing. Several gluings give rise to identical interference
phenomena. Due to these limitations we here introduce a generalized
interferometer. It is shown that this setup has the capacity to
distinguish all possible trace preserving gluings of arbitrary
channels. As such this provides a tool for experimental investigations
of which gluings are present in actual evolutions.

\begin{acknowledgments}
I thank Erik Sj\"oqvist for many valuable comments and discussions.
\end{acknowledgments}

\end{document}